\documentclass[12pt]{article}
\pdfoutput=1
\usepackage{amssymb,amsmath,bbold}
\usepackage{mathrsfs}
\usepackage{dsfont}
\usepackage{graphicx}
\usepackage{hyperref}
\usepackage{cite}
\numberwithin{equation}{section}

\def\singleandhalfspaced{\baselineskip=\normalbaselineskip\multiply
    \baselineskip by 150\divide\baselineskip by 100}

\newenvironment{Eqnarray}%
         {\arraycolsep 0.14em\begin{eqnarray}}{\end{eqnarray}}
 \newcommand{\be}{\begin{Eqnarray}} 
  \newcommand{\ee}{\end{Eqnarray}}

\let\Re\relax
\let\Im\relax
\DeclareMathOperator{\Re}{Re}
\DeclareMathOperator{\Im}{Im}

\def\iso{\mathchoice{\cong}{\cong}{\isoS}{\cong}}
\def\isoS{\vbox{\baselineskip 0pt  \lineskip 0.5pt
    \ialign{$ \mathsurround=0pt  \scriptstyle \hfil ## \hfil $\crcr
        \sim \crcr = \crcr}}}
\newcommand{\beq}{\begin{equation}}
\newcommand{\eeq}{\end{equation}}
\newcommand{\beqa}{\begin{Eqnarray}}
\newcommand{\eeqa}{\end{Eqnarray}}
\def\vev#1{\langle #1 \rangle}

\def\eq#1{Eq.~(\ref{#1})}

\def\eqs#1#2{Eqs.~(\ref{#1}) and (\ref{#2})}

\def\eqst#1#2{Eqs.~(\ref{#1})--(\ref{#2})}

\def\ifmath#1{\relax\ifmmode #1\else $#1$\fi}
\def\lsim{{~\raise.15em\hbox{$<$}\kern-.85em\lower.35em\hbox{$\sim$}~}}
\def\gsim{{~\raise.15em\hbox{$>$}\kern-.85em\lower.35em\hbox{$\sim$}~}}
\def\ls#1{\ifmath{_{\lower1.5pt\hbox{$\scriptstyle #1$}}}}

\def\half{\tfrac{1}{2}}

\title{
\vspace*{-2.3cm}
\begin{flushright}
\normalsize{
SCIPP-16/06 \\
NSF-KITP-16-048
}
\end{flushright}
\vspace{1.5cm}
\Large
\textbf{
Partially Natural Two Higgs Doublet Models
\\
}\vspace*{1.0cm}
}

\author{Patrick Draper$^1$, Howard E.~Haber$^{2,3}$, and Joshua T.~Ruderman$^4$
\vspace{5mm}
\\
\normalsize\emph{$^1$Department of Physics, University of California, Santa Barbara, CA 93106} \\
\normalsize\emph{$^2$Santa Cruz Institute for Particle Physics, University of California, Santa Cruz CA 95064} \\
\normalsize\emph{$^3$Kavli Institute for Theoretical Physics, University of California, Santa Barbara, CA 93106}\\
\normalsize\emph{$^4$Center for Cosmology and Particle Physics},\\[-1mm] \normalsize \emph{Department of Physics, New York University, New York, NY 10003}
}

\begin{document}
\singleandhalfspaced
\setcounter{page}{0}
\date{\vspace{-5ex}}
\maketitle

\vspace*{1cm}
\begin{abstract}

It is possible that the electroweak scale is low due to the fine-tuning of microscopic parameters, which can result from selection effects.
The experimental discovery of new light fundamental scalars other than the Standard Model Higgs boson would seem to disfavor this possibility, since generically such states imply parametrically worse fine-tuning with no compelling connection to selection effects. 
 We discuss counterexamples where the Higgs boson is light because of fine-tuning, and a second scalar doublet is light because a discrete symmetry relates its mass to the mass of the Standard Model Higgs boson. 
Our examples require new vectorlike fermions at the electroweak scale, and the models possess a rich electroweak vacuum structure.  The mechanism that we discuss does not protect a small CP-odd Higgs mass in split or high-scale supersymmetry-breaking scenarios of the MSSM due to an incompatibility between the discrete symmetries and holomorphy.

\end{abstract}

\thispagestyle{empty}
\newpage

\vspace{-3cm}

\setcounter{footnote}{0} \setcounter{page}{2}
\setcounter{section}{0} \setcounter{subsection}{0}
\setcounter{subsubsection}{0}


\section{Introduction and Conclusions}

The hierarchy between the electroweak scale and a high scale cutoff, $\Lambda$, may result from a finely-tuned Higgs mass.  For example, $\Lambda$ may represent the Planck scale, the grand unification (GUT) scale\cite{Raby:2008zz}, or a high supersymmetry (SUSY) breaking scale\cite{Haber:2008zz}.  One possible mechanism for producing a tuned electroweak scale is environmental selection in a multiverse~\cite{Agrawal:1997gf,Damour:2007uv,Hall:2014dfa}, in analogy to Weinberg's anthropic constraint on the size of the cosmological constant~\cite{Weinberg:1987dv}.  In a {\it generic} theory, additional tunings are required if additional light states exist in the scalar spectrum.  Since only one light scalar Higgs boson is required to break electroweak symmetry,  it is reasonable to expect that even if one Higgs state is tuned to be light, $m_h \ll \Lambda$, all other scalars still reside near the cutoff scale $\Lambda$.  This reasoning suggests that the infrared physics should consist of the Standard Model, possibly augmented by new fermions, whose masses are protected by chiral symmetry, but not new scalars.
Examples of this class of theories are the Standard Model (SM) extrapolated to high energies, Split Supersymmetry~\cite{ArkaniHamed:2004fb}, and the SM plus fermionic dark matter~\cite{Cirelli:2005uq}.

Sometimes, the above argument is turned around.  If new fundamental scalars are discovered at the LHC, then selection probably does not explain the electroweak scale~\cite{ArkaniHamed:2012kq}.  This points to a natural electroweak scale, perhaps with supersymmetry around the corner. More radically, extra scalars and the absence of a stabilizing symmetry may be evidence 
that quantum gravity does not contribute a large correction to scalar masses~\cite{Dubovsky:2013ira}.
However, while compelling for generic theories, this conclusion is not a theorem, and can fail for non-generic theories. 
A simple counterexample is the case where the masses of one more more new scalars are linked 
by symmetry to the mass of the SM Higgs boson.  In such theories, only one tuning is required, and results in additional light scalars.\footnote{In the multiverse scenario, this requires typical vacua satisfying selection criteria to possess extra states and symmetries in the Higgs sector.  This is possible, given our ignorance of the underlying distribution of vacua.  Note that we already observe  our vacuum to contain more particles and symmetries than seem necessary for observers, such as three fermion generations with approximate flavor symmetries.}

In this article we discuss a simple Two Higgs Doublet Model (2HDM)\cite{Branco:2011iw}
 of this type. Discrete symmetries constrain the mass terms in the Higgs sector so that when one doublet is tuned light, the other comes along for the ride. The symmetries also constrain the quartic couplings, and in the Yukawa sector they require the introduction of new fermion states with soft symmetry breaking vectorlike mass terms. We analyze the Higgs sector phenomenology and the naturalness constraints on the symmetry breaking terms. 

Similar 2HDMs were previously studied in~\cite{Pierce:2007ut}, although with a different motivation.  Ref.~\cite{Pierce:2007ut} set out to identify models where the Higgs mass is fine-tuned, but related to the dark matter mass by a symmetry, preserving weak scale dark matter.  They studied the Inert Double Model~\cite{Ma:2006km,Barbieri:2006dq}, where an unbroken $\mathbb{Z}_2$ protects the stability of one doublet, which is identified with dark matter.  The dark matter mass is related to the Higgs mass by a discrete symmetry, which requires vectorlike fermions.  Our focus will be on a similar model with a slightly different vectorlike fermion content. We present a general classification of the symmetries that can tie the two masses and identify the minimal requirements. We analyze the full vacuum structure, allowing the inert symmetry to be spontaneously broken, and we provide updated constraints given LHC measurements of the Higgs mass and couplings. 

We then discuss the extension of these symmetries to supersymmetric models. Given our minimal example of a ``partially natural" 2HDM, it is interesting to think how it may be embedded in more a ultraviolet (UV) complete 2HDM\@. For example, if the soft SUSY breaking scale in the MSSM is high, the ``generic theory" expectation is that the mass of the second Higgs doublet is of order $m_{soft}$ -- far in the decoupling limit. Can the discrete symmetries be implemented so that the mass of the CP-odd Higgs boson, $m_A$, is naturally of order the electroweak scale? We show that the mechanism discussed in the non-supersymmetric case is not extendable in a simple way. In supersymmetry, some of the discrete symmetries take the form of generalized charge conjugation or parity symmetries that cannot be extended to the Yukawa and gauge sectors simultaneously, inevitably introducing hard breaking of the symmetries. However, in the MSSM without extra matter, we find that there are corners of parameter space where there is a hierarchy between $m_A$ and $m_{soft}$ and hard discrete symmetry breaking effects on the fine-tuning are relatively mild. 

The remainder of this paper is organized as follows.  In section~\ref{sec:finetuned}, we study all possible symmetries of the 2HDM and determine when one fine-tuning implies more than one light scalar.  In section~\ref{sec:higgs}, we specialize to a specific example, with a $\mathbb{Z}_2 \times \mathbb{Z}_2$ symmetry, which we study in the remainder of the paper.  In section~\ref{sec:Yukawa}, we extend the discrete symmetry to the  Yukawa sector, which requires adding vectorlike fermions whose masses softly break the symmetry.  We discuss discrete symmetry breaking effects in section~\ref{sec:BreakDiscrete}, the vacuum structure in section~\ref{sec:vacua}, the scalar spectrum in section~\ref{sec:spec}, and the phenomenology of the vectorlike quarks in section~\ref{sec:pheno}.  In section~\ref{sec:SUSY}, we show that the discrete symmetry cannot be extended in a simple way to protect the mass of the heavy doublet in the MSSM\@. However, in gauge-mediated models with large $\tan\beta$, an approximate discrete symmetry between the top and bottom Yukawa couplings may allow a natural one-loop splitting between the second doublet and the soft SUSY-breaking scale.

\section{Fine-Tuning of Parameters in the Two Higgs Doublet Model}
\label{sec:finetuned}

The Higgs sector of the Standard Model is famously unnatural~\cite{Weisskopf:1939zz,Altarelli:2013lla}.\footnote{It has been suggested~\cite{Grzadkowski:2009iz} that a general Higgs sector may be natural if the couplings satisfy Veltman conditions~\cite{Veltman:1980mj}. However, such a non-generic model is natural only if the values of couplings compatible with the Veltman conditions also lie near an infrared fixed point, and any viable model of this type is likely to require a considerable amount of extra structure near the weak scale~\cite{Houtz:2016jxk}.}   In particular, a fine-tuning of the Higgs sector squared-mass parameter is required in order to explain the observed vacuum expectation value of the neutral Higgs field: $v\simeq 246$~GeV\@.    
As we now demonstrate, the Two-Higgs Doublet Model (2HDM) generically requires 4 separate and independent fine-tunings.  To determine the
number of independent fine-tunings consider the most general 
renormalizable SU(2)$_L\times$U(1)$_Y$ invariant 2HDM scalar potential (see e.g., Ref.~\cite{Gunion:2002zf}),
\be
\mathcal{V}&=& m_{11}^2\Phi_1^\dagger\Phi_1+m_{22}^2\Phi_2^\dagger\Phi_2
-[m_{12}^2\Phi_1^\dagger\Phi_2+{\rm h.c.}]
 +\half\lambda_1(\Phi_1^\dagger\Phi_1)^2
+\half\lambda_2(\Phi_2^\dagger\Phi_2)^2
+\lambda_3(\Phi_1^\dagger\Phi_1)(\Phi_2^\dagger\Phi_2)\nonumber\\[8pt]
&&\qquad\qquad\,\,
+\lambda_4(\Phi_1^\dagger\Phi_2)(\Phi_2^\dagger\Phi_1)
+\left\{\half\lambda_5(\Phi_1^\dagger\Phi_2)^2
+\big[\lambda_6(\Phi_1^\dagger\Phi_1)
+\lambda_7(\Phi_2^\dagger\Phi_2)\big]
\Phi_1^\dagger\Phi_2+{\rm h.c.}\right\}, \label{pot}
\ee 
where $\Phi_1$ and $\Phi_2$ (which defines the \textit{generic basis} of fields) denote two complex $Y=1$, SU(2)$\ls{L}$ doublet scalar fields.  Note that the parameters $m_{11}^2$, $m_{22}^2$ and $\lambda_{1,2,3,4}$ are real, whereas the parameters $m_{12}^2$ and $\lambda_{5,6,7}$ are potentially complex.
When the scalar potential is minimized, the neutral components of $\Phi_1$ and~$\Phi_2$ acquire vacuum expectation values [vevs],
$\langle\Phi_i^0\rangle=v_i/\sqrt{2}$ where  $v_i\equiv |v_i| e^{i\xi_i}$ and 
$v^2\equiv |v_1|^2+|v_2|^2$.  
It is conventional to write
\beq \label{betadef}
c_\beta\equiv \cos\beta=|v_1|/v\,,\qquad\quad s_\beta\equiv \sin\beta=|v_2|/v\,,
\eeq
where $0\leq\beta\leq\half\pi$.  Furthermore, we define the relative phase of the vevs, $\xi\equiv \xi_2-\xi_1$.

Since $\Phi_1$ and $\Phi_2$ are indistinguishable fields,
it is always possible to define two orthonormal linear combinations of the two
doublet fields without modifying any prediction of the model.  In particular, the \textit{Higgs basis}
is obtained by defining new Higgs doublet fields,
\be
H_1=\begin{pmatrix}H_1^+\\ H_1^0\end{pmatrix}\equiv \frac{v_1^* \Phi_1+v_2^*\Phi_2}{v}\,,
\qquad\quad H_2=\begin{pmatrix} H_2^+\\ H_2^0\end{pmatrix}\equiv\frac{-v_2 \Phi_1+v_1\Phi_2}{v}
 \,,
\ee
where $\vev{H_1^0}=v/\sqrt{2}$, $\vev{H_2^0}=0$, and $v=246$~GeV\@.

In terms of the Higgs basis fields, the scalar potential takes on a form similar to \eq{pot}~\cite{Davidson:2005cw},
\be \!\!\!\!\!\!\!\!
 \mathcal{V} &=& Y_1 H_1^\dagger H_1+Y_2 H_2^\dagger H_2
   +\left(Y_3 H_1^\dagger H_2+{\rm h.c.}\right) +\half Z_1(H_1^\dagger H_1)^2
   +\half Z_2(H_2^\dagger H_2)^2+Z_3 (H_1^\dagger H_1) (H_2^\dagger H_2)  \nonumber \\
&& \qquad\qquad
   +Z_4 (H_1^\dagger H_2)(H_2^\dagger H_1) 
   +\left[\half Z_5(H_1^\dagger H_2)^2+[Z_6 H_1^\dagger H_1+Z_7 H_2^\dagger H_2]H_1^\dagger H_2+{\rm h.c.}\right]\!,
\ee
where the $Y_i$ are linear combinations of the squared-mass parameters, $m^2_{ij}$ [cf.~\eqst{y1}{y3}], and the $Z_i$ are linear combinations of the $\lambda_i$.   In general, the parameters $Y_{1,2}$ and $Z_{1,2,3,4}$ are real and the parameters $Y_3$, $Z_{5,6,7}$ are potentially complex.
Assuming that 
$|\lambda_i|/(4\pi)\lsim\mathcal{O}(1)$ (so that the scalar potential satisfies unitarity constraints\cite{Kanemura:2015ska} at tree-level), it follows that the $Z_i$ cannot become arbitrarily large.     
The scalar potential minimum conditions in the Higgs basis are:
\beq \label{yone}
Y_1=-\tfrac{1}{2}Z_1 v^2\,,\qquad\quad Y_3=-\tfrac{1}{2} Z_6 v^2\,.
\eeq
Note that the most general 2HDM is governed by 11 free parameters.  The counting is most easily done in the Higgs basis, where in light of \eq{yone} we identify the independent parameters to be the vev $v$, $Y_2$, $Z_{1,2,3,4}$, $|Z_{5,6,7}|$ and two relative phases among the complex parameters $Z_{5,6,7}$.  

We now turn to the number of fine-tuning conditions required in order that all Higgs masses are of $\mathcal{O}(v)$.  We need one fine-tuning condition to set the scale of electroweak symmetry breaking to be $v=246$~GeV (rather than some high energy energy scale $\Lambda\gg v$).  This is equivalent to fine-tuning the value of $Y_1\sim\mathcal{O}(v^2)$ in \eq{yone}, under the assumption that $Z_1\sim\mathcal{O}(1)$.\footnote{Indeed, in the alignment limit where one of the neutral Higgs states $h$ exhibits couplings that approximate those of the Standard Model (SM) Higgs boson, 
we have to good approximation $m_h^2\simeq \half Z_1 v^2$, which yields $Z_1\simeq 0.26$.}  Likewise, 
under the assumption that $Z_6$ is an $\mathcal{O}(1)$ parameter, 
it also follows from \eq{yone} that  $Y_3\sim\mathcal{O}(v^2)$.  However, the parameter~$Y_2$, which sets the 
squared-mass scale of the non-SM-like Higgs states, is not yet fixed.   One can easily derive the following squared-mass relations,
\beq
m_{H^\pm}^2=Y_{2}+\half Z_3 v^2\,,\qquad\quad {\rm Tr}~\mathcal{M}^2_H=2Y_2+(Z_1+Z_3+Z_4)v^2\,,
\eeq
where $\mathcal{M}^2_H$ is the $3\times 3$ neutral Higgs squared-mass matrix.  Thus, to ensure all non-SM-like Higgs boson masses of $\mathcal{O}(v)$, one must perform additional fine-tunings in order to set $Y_2\sim\mathcal{O}(v^2)$.

The total number additional fine-tunings required depends on how the parameters of the model are scanned. The natural values for the squared-mass parameters $m_{11}^2$, $m_{22}^2$ and $|m_{12}^2|$ of the scalar potential defined in \eq{pot} are of $\mathcal{O}(\Lambda^2)$, where $\Lambda$ is the ultraviolet cutoff of the theory.   Fine-tuning conditions correspond to the requirement that certain linear combinations of these parameters must yield squared-mass parameters of $\mathcal{O}(v^2)$.  To see how this works in practice in the case of the 2HDM, we write the squared-mass parameters of the scalar potential in the Higgs basis, $Y_1$, $Y_2$ and $Y_3$, in terms of $m_{11}^2$, $m_{22}^2$ and $m_{12}^2$~\cite{Davidson:2005cw},
\beqa
Y_1&=&-\half Z_1 v^2=m_{11}^2 c^2_\beta+m_{22}^2 s^2_\beta- \Re(m_{12}^2 e^{i\xi})s_{2\beta}\,,\label{y1} \\
Y_2&=& m_{11}^2 s^2_\beta+m_{22}^2 c^2_\beta+\Re(m_{12}^2 e^{i\xi})s_{2\beta}\,,\label{y2} \\
Y_3&=&-\half Z_6 v^2 =-e^{-i\xi}\bigl[\half(m_{11}^2-m_{22}^2)s_{2\beta}+\Re(m_{12}^2 e^{i\xi})c_{2\beta}+i\Im(m_{12}^2 e^{i\xi})\bigr]\,,\label{y3}
\eeqa
where $s_{2\beta}\equiv \sin 2\beta$, $c_{2\beta}\equiv \cos 2\beta$, and $\xi$ is defined below \eq{betadef}.  Since $Y_3$  and $m_{12}^2$ are generically complex, \eqst{y1}{y3} constitute four real equations.   Two of these equations can be used to fix the values of $\beta$ and $\xi$ (which determine the vevs $v_i$ up to an overall scale).  The two remaining equations can be used to fix the value of $v=246$~GeV and to set $Y_2\sim\mathcal{O}(v^2)$.  The first of these two equations is obtained by adding \eqs{y1}{y2}.  The second of the two equations is obtained by eliminating $\beta$
by combining \eqs{y1}{y3}.  Explicitly,  
\beqa
m_{11}^2+m_{22}^2&=&Y_2-\half Z_1 v^2\,, \label{ft1} \\
(m_{11}^2-m_{22}^2)^2+4|m_{12}^2|^2&=&(Y_2+\half Z_1 v^2)^2+|Z_6 v^2|^2\,. \label{ft2}
\eeqa  
Hence, $m_{11}^2$, $m_{22}^2$ and $|m_{12}^2|$ must \textit{all} be of $\mathcal{O}(v^2)$ in order to set $Y_2\sim\mathcal{O}(v^2)$.  Since the natural values of $m_{11}^2$, $m_{22}^2$, $\Re(m_{12}^2)$ and $\Im(m_{12}^2)$ are of $\mathcal{O}(\Lambda^2)$, \eqs{ft1}{ft2} correspond to \textit{four} fine-tuning conditions.\footnote{The total number of tunings depends on how $m_{12}^2$ is parameterized ({\it i.e.}~how $m_{12}^2$ scans).  If $m_{12}^2$ is parameterized by $\Re(m_{12}^2)$ and $\Im(m_{12}^2)$, then \eq{ft2} implies 4 tunings, whereas if $m_{12}^2$ is parameterized by $|m_{12}^2|$ and a phase, then there are 3 tunings since \eq{ft2} depends only on $|m_{12}^2|$.}  In particular, \eq{ft1} yields one fine-tuning condition whereas \eq{ft2} yields three fine-tuning conditions since the left-hand side of \eq{ft2} is the sum of squares, which implies that $m_{11}^2-m_{22}^2$, $\Re(m_{12}^2)$ and $\Im(m_{12}^2)$ are separately of $\mathcal{O}(v^2)$.

Note that in the decoupling limit of the 2HDM where $Y_2\gg v^2$, we can combine \eqs{ft1}{ft2} to obtain 
\beq \label{ft3}
|m_{12}^2|^2-m_{11}^2 m_{22}^2=\half v^2\bigl[Z_1 Y_2+\half|Z_6|^2 v^2\bigr]\,.
\eeq
If $Y_2\sim\mathcal{O}(\Lambda^2)$, then it is possible to have $m_{11}^2$, $m_{22}^2$, $\Re(m_{12}^2)$ and $\Im(m_{12}^2)$ all of $\mathcal{O}(\Lambda^2)$, with one fine-tuning condition given by \eq{ft3} to fix $v\ll \Lambda$.  This is equivalent to the usual fine-tuning condition of the SM Higgs sector to fix the scale of electroweak symmetry breaking.

\begin{table}[t!]
\begin{tabular}{ccccccccccccc}
symmetry &  $m_{11}^2$ & $m_{22}^2$ & $m_{12}^2$ & $\lambda_1$ &
 $\lambda_2$ & $\lambda_3$ & $\lambda_4$ &
$\lambda_5$ & $\lambda_6$ & $\lambda_7$ \\
\hline
$\mathbb{Z}_2^i$ & &   & 0
   &  &  &  & &
   & 0 & 0 \\
$\mathbb{Z}_2^m$  &  &$ m_{11}^2$ & real &&
    $ \lambda_1$ & &  &  
real &  & $\lambda_6^\ast$
\\
U(1) & &  & 0 
 &  & &  & &
0 & 0 & 0 \\
SO(3)  && $ m_{11}^2$ & 0
   && $\lambda_1$ &  & $\lambda_1 - \lambda_3$ &
0 & 0 & 0 \\
CP1   & & & real
 & &  &  &&
real & real & real \\
CP2   && $m_{11}^2$ & 0 
  && $\lambda_1$ &  &  &
   &  & $- \lambda_6$ \\
CP3   && $m_{11}^2$ & 0 
   && $\lambda_1$ &  &  &
$\lambda_1 - \lambda_3 - \lambda_4$ (real) & 0 & 0 \\
\end{tabular}
\caption{Classification of 2HDM scalar potential symmetries and their impact on the coefficients of the scalar potential [cf.~\eq{pot}] in a generic basis~\cite{Ivanov:2007de,Ferreira:2009wh,Ferreira:2010yh,Battye:2011jj}.
Empty entries in Table~\ref{tab:class} correspond to a lack of constraints on the corresponding parameters. Strictly speaking, the $\mathbb{Z}_2^m$ is not an independent symmetry condition, since a change of scalar field basis can be performed in this case to a new basis in which the $\mathbb{Z}_2^i$ symmetry is manifest.
\label{tab:class}}
\end{table}

We propose to remove the additional fine-tuning conditions by imposing a discrete or continuous symmetry on the scalar potential given in \eq{pot}, which reduces the number of independent parameters.
Remarkably, the possible symmetries that yield distinct models are quite limited and are listed in Table~1~\cite{Ivanov:2007de,Ferreira:2009wh,Ferreira:2010yh,Battye:2011jj}.
These symmetries fall in two classes:
Higgs family symmetries and generalized CP symmetries.    
The Higgs family symmetry transformations are subgroups of the U(2) transformation, $\Phi_a\to U_{ab}\Phi_b$, where $U$ is a $2\times 2$ unitary matrix.  The U(1)$_Y$ hypercharge symmetry group (corresponding to  $\Phi_a\to e^{i\alpha}\Phi_a$) is a subgroup of U(2) that is always present due to electroweak gauge invariance.   The generalized CP (or GCP) symmetries are generically of the form $\Phi_a\to U_{ab}\Phi_b^*$.
In the list of Higgs family and GCP symmetries given below and in Table~\ref{tab:class}, an element of the corresponding symmetry group should be regarded as an equivalence class of elements, where the elements of a given class are related by U(1)$_Y$ symmetry transformations.\footnote{A larger class of accidental symmetries of the 2HDM scalar potential that utilizes mixed Higgs family and generalized CP symmetries, which has been treated in Refs.~\cite{Battye:2011jj,Pilaftsis:2011ed,Dev:2014yca}, will not be considered here.}

We summarize below the symmetry transformations of the scalar fields corresponding to the symmetries listed in Table~\ref{tab:class}.  The possible Higgs family symmetries modulo the
U(1)$_Y$ hypercharge symmetry can be either discrete or continuous:

$\mathbb{Z}_2^i:\hspace{43ex}
\Phi_1 \rightarrow -\Phi_1,
\hspace{7.5ex}
\Phi_2 \rightarrow  \Phi_2$

$\mathbb{Z}_2^m$\,\,\, ({\rm mirror symmetry}):
\hspace{21.2ex}
$\Phi_1 \longleftrightarrow \Phi_2$

U(1)$_{\rm PQ}$\,\,\,({\rm Peccei-Quinn symmetry~\cite{Peccei:1977hh}}):
\hspace{7.5ex}\!
$\Phi_1 \rightarrow e^{-i \theta} \Phi_1$,
\hspace{4ex}
$\Phi_2 \rightarrow e^{i \theta} \Phi_2$

SO(3):\,\,\,({\rm maximal Higgs flavor symmetry)}:\hspace{5.5ex} $\!\Phi_a\to U_{ab}\Phi_b$\,,\qquad 
\,\, $U\in{\rm U(2)}/{\rm U(1)}_Y$

\noindent 
It should be noted that starting from the scalar potential of a $\mathbb{Z}_2^m$-symmetric 2HDM, one can find a different basis of scalar fields in which the corresponding scalar potential manifestly exhibits the $\mathbb{Z}_2^i$ symmetry, and vice versa~\cite{Davidson:2005cw}.

The GCP symmetry transformations of the scalar fields listed in Table~\ref{tab:class} are as follows:

${\rm CP1}:\hspace{6ex}
\Phi_1 \rightarrow \Phi_1^*,
\hspace{20.5ex}
\Phi_2 \rightarrow \Phi_2^*$

${\rm CP2}:\hspace{6ex}
\Phi_1 \rightarrow \Phi_2^*,
\hspace{20.5ex}
\Phi_2 \rightarrow -\Phi_1^*$

${\rm CP3}:\hspace{6ex}
\Phi_1 \rightarrow \Phi_1^*\cos\theta+\Phi_2^*\sin\theta,
\hspace{6ex}\!
\Phi_2 \rightarrow -\Phi_1^*\sin\theta+\Phi_2^*\cos\theta,\qquad \text{for $0<\theta<\half\pi$}$

\noindent
In all cases, each of the above symmetries is applied singly in a generic $\{\Phi_1,\Phi_2\}$ basis.
One can also consider the possibility of applying two of the symmetries listed above simultaneously in the same basis.  It was shown in Ref.~\cite{Ferreira:2009wh} that no new independent models arise in this way.  For example, applying $\mathbb{Z}_2^i$ and $\mathbb{Z}_2^m$ in the same basis yields a $\mathbb{Z}^i_2\otimes \mathbb{Z}_2^m$ model that is equivalent to CP2 when expressed in a different basis.\footnote{Note that the discrete $\mathbb{Z}^i_2$ and $\mathbb{Z}_2^m$ symmetry transformations commute modulo the hypercharge U(1)$_Y$ group.  
That is, if we define the subgroup $Z_Y=\{\mathds{1}\,,\,-\mathds{1}\}$ of U(1)$_Y$ (where $\mathds{1}$ is the $2\times 2$ identity matrix), then the elements of $\mathbb{Z}^i_2=\{\mathds{1}\,,\,\sigma_3\}$, $ \mathbb{Z}_2^m=\{\mathds{1}\,,\,\sigma_1\}$ and $Z_Y$ generate the dihedral group $D_4=\{\pm\mathds{1}\,,\,\pm\sigma_1\,,\,\pm i\sigma_2\,,\,\pm\sigma_3\}$ of eight 
elements~\cite{Ferreira:2009wh,Pierce:2007ut}
 (here, the $\sigma_i$ are the usual Pauli matrices).  However, if we impose an equivalence relation by identifying $g$ and $-g$ for all $g\in D_4$, then we recover the group of four elements,
$D_4/Z_Y\iso D_2\iso \mathbb{Z}^i_2\otimes \mathbb{Z}_2^m$.}
Similarly, applying U(1)$_{\rm PQ}$ 
and $\mathbb{Z}_2^m$ in the same basis yields a model that is equivalent to CP3 when expressed in a different basis.  Other examples of these types can be found in Ref.~\cite{Ferreira:2009wh}.  

One can automatically remove one of the fine-tuning conditions of the 2HDM by choosing a CP-invariant scalar potential, which sets $\Im(m_{12}^2)=0$ in a real basis.  By imposing any of the other 
symmetries of Table~\ref{tab:class}, one can remove either two or three fine-tuning conditions depending on the choice of symmetry.  For example, imposing a
$Z_2^i$ or U(1)$_{\rm PQ}$ symmetry on the scalar potential sets the complex parameter $m_{12}^2=0$, thereby removing two fine-tuning conditions.  In order to remove three fine-tuning conditions (leaving only the fine-tuning condition associated with setting $v=246$~GeV), we shall impose a
symmetry on the scalar potential that places the 2HDM in an exceptional region of parameter space, first identified in Ref.~\cite{Davidson:2005cw}  and later dubbed the ERPS in Ref.~\cite{Ferreira:2009wh}.  This parameter regime consists of choosing
\beq
m_{22}^2=m_{11}^2\,,\qquad m_{12}^2=0\,,\qquad \lambda_1=\lambda_2\,,\qquad \lambda_7=-\lambda_6\,.\label{erpsgen}
\eeq
The corresponding conditions in the Higgs basis are,
\beq \label{erps}
Y_2=Y_1\,,\qquad Y_3=Z_6=Z_7=0\,,\qquad Z_1=Z_2\,.
\eeq
Indeed in the ERPS, $Y_2$ is no longer an independent parameter, which means that the fine-tuning conditions that set $Y_2\sim\mathcal{O}(v^2)$ are automatically satisfied once the first fine-tuning condition to set $Y_1\sim\mathcal{O}(v^2)$ is implemented.

The ERPS is consistent with three of the symmetries listed in Table~\ref{tab:class}: SO(3), CP2 (equivalent to $\mathbb{Z}^i_2\otimes \mathbb{Z}_2^m$ in another basis) and CP3 (equivalent to U(1)$\otimes \mathbb{Z}_2^m$ in another basis).  The 2HDMs that are constrained either by an SO(3) or CP3 symmetry contain an extra massless (CP-odd) Goldstone boson due to an underlying Peccei-Quinn symmetry (this state can be lifted to a pseudo-Goldstone boson by softly breaking the global symmetry).  In contrast, in the CP2-symmetric 2HDM, the physical CP-odd Higgs boson is massive.  We shall henceforth focus on the CP2-symmetric 2HDM as a potential model of an extended Higgs sector with one fine-tuning condition.

Until now, we have focused solely on the pure scalar sector of the 2HDM\@.  Including the interactions with the vector bosons does not modify the analysis above, since these arise from the scalar kinetic energy terms when an ordinary derivative is replaced by a gauge covariant derivative.  The resulting kinetic energy term is in fact invariant under the full U(2) Higgs family symmetry.  However, when we include the Yukawa couplings of the Higgs bosons to the fermions, we immediately face a challenge.
The Yukawa interactions are linear in the scalar fields that transform non-trivially with respect to the Higgs family and GCP symmetries.  Thus, if we want to preserve the symmetry that yields \eq{erps} and preserves the single fine-tuning, then we must introduce transformation laws for the fermion fields in such a way that respect the corresponding Higgs family or GCP symmetry.   This in turn will constrain the structure of the Higgs-fermion interactions. 

Such an analysis was performed in Ref.~\cite{Ferreira:2010bm}, where the following results were obtained.  First, it was shown that there is no extension of the CP2 symmetry to the Higgs--quark Yukawa couplings in a way consistent with experiment.  In particular, a CP2-symmetric Higgs Lagrangian necessarily contains a massless quark.
Second, all possible extensions of the CP3 symmetry (corresponding to different choices of $\theta$) were considered.  Only one potentially viable CP3-symmetric 2HDM was found, although this model yielded a value of the Jarlskog invariant that was nearly three orders of magnitude below the experimentally observed value.  Moreover, this model possesses tree-level Higgs-mediated flavor-changing neutral currents that may already be inconsistent with data.  Finally, as previously noted, an exact CP3 symmetry yields a massless CP-odd scalar and thus must be softly-broken by taking $m_{12}^2\neq 0$ in \eq{erpsgen}.  However, the latter would introduce additional fine-tuning if $m_{12}^2$ were much larger than the electroweak scale. 

In order to extend the CP2 symmetry of the bosonic part of the 2HDM Lagrangian to the full Higgs Lagrangian, we will need to add additional fermions to the model.  This in turn will allow us to modify the transformation laws of the ordinary fermions under CP2 and avoid the conclusions of 
Ref.~\cite{Ferreira:2010bm}.

\section{Model of a partially natural 2HDM}
\label{sec:model}

\subsection{Higgs Sector}
\label{sec:higgs}
We focus on the CP2-symmetric 2HDM scalar potential, which was shown in Section~\ref{sec:finetuned} to alleviate two of the three fine-tuning conditions of the model.  As noted above, the CP2 model is equivalent to imposing simultaneously the discrete $\mathbb{Z}_2^m$ and $\mathbb{Z}_2^i$ symmetries on the scalar potential.  

It is instructive to exhibit how the second fine-tuning is alleviated in the generic $\{\Phi_1,\Phi_2\}$ basis.  Consider the mass terms of the general 2HDM,
\be
V \, \supset \,  m_{11}^2 \Phi_1^\dagger \Phi_1 + m_{22}^2 \Phi_2^\dagger \Phi_2- \left[m_{12}^2 \Phi_1^\dagger \Phi_2 + \rm{h.c.} \right].
\ee
In order to relate the two scalars by a symmetry, we introduce a $\mathbb{Z}_2$ exchange, or mirror\footnote{Here, we restrict to a single copy of the SM gauge group, $G_{SM}=SU(3)_c\times SU(2)_W \times U(1)_Y$, unlike many studies (see for example Refs.~\cite{Foot:1991bp, Barbieri:2005ri}) that consider a mirror symmetry that exchanges two copies of the SM gauge group.}, symmetry,
\be\label{eq:mirrorsym}
\mathbb{Z}_2^m: \quad \Phi_1 \Longleftrightarrow \Phi_2.
\ee
The mirror symmetry sets $m \equiv m_{11} = m_{22}$,
\be
V \, \supset \,  m^2 \left( \Phi_1^\dagger \Phi_1 +  \Phi_2^\dagger \Phi_2\right)- \left[m_{12}^2 \Phi_1^\dagger \Phi_2 + \rm{h.c.} \right].
\ee
This mirror symmetry is insufficient to ensure two light Higgs bosons at low energies because the eigenvalues of the mass matrix are given by $m^2 \pm m_{12}^2$.  This means that one Higgs boson can be tuned light, with one heavy, by setting $m^2 \approx m_{12}^2$ with both parameters of order $\Lambda^2$.  Another way to see why the mirror symmetry alone is insufficient is to change variables, $\Phi_\pm = \Phi_2 \pm \Phi_1$.  In these variables $\mathbb{Z}_2^m$ acts as,
\be
\mathbb{Z}_2^m: \quad \Phi_+ \Longleftrightarrow \Phi_+, \quad \Phi_- \Longleftrightarrow - \Phi_-.
\ee
In particular,
$\Phi_+$ and $\Phi_-$ transform differently and have independent masses.

In order to ensure two light eigenvalues we need to introduce a second symmetry where $\Phi_1^\dagger \Phi_2$ transforms non-trivially such that $m_{12}^2$ is forbidden.  A simple possibility is to introduce a second $\mathbb{Z}_2$ symmetry where $\Phi_1$ is odd and $\Phi_2$ is even,
\be
\mathbb{Z}_2^i: \quad \Phi_1 \Longleftrightarrow - \Phi_1, \quad \Phi_2 \Longleftrightarrow  \Phi_2.
\ee
The ``i" stands for ``inert, " since in the inert phase $\langle \Phi_1\rangle=0$ and thus $\mathbb{Z}_2^i$ is unbroken. Now, both scalars have mass $m^2$.  In the $\Phi_\pm$ variables, $\mathbb{Z}_2^i$ acts as an {\it exchange} symmetry, $\Phi_+ \Longleftrightarrow \Phi_-$.  The full $\mathbb{Z}_2^m \times\mathbb{Z}_2^i$ symmetry is necessary to guarantee that an exchange symmetry is present independent of field redefinitions. 

In the general 2HDM the quartic interactions are given by~\eq{pot},
where $\lambda_5, \lambda_6$, and $\lambda_7$ are in general complex.  The mirror symmetry, $\mathbb{Z}_2^m$, requires that $\lambda \equiv \lambda_1 = \lambda_2$, $\lambda_5$ is real, and $\lambda_6 = \lambda_7^*$.  The $\mathbb{Z}_2^i$ symmetry sets $\lambda_6 = \lambda_7 = 0$.  
To summarize, the 2HDM potential invariant under $\mathbb{Z}_2^m \times\mathbb{Z}_2^i$ is given by
\be 
V \, &\supset &  m^2 \left( \Phi_1^\dagger \Phi_1 +  \Phi_2^\dagger \Phi_2\right) 
 +  \tfrac{1}{2} \lambda \left[  (\Phi_1^\dagger \Phi_1)^2 + (\Phi_2^\dagger \Phi_2)^2 \right] +\lambda_3 (\Phi_1^\dagger \Phi_1)(\Phi_2^\dagger \Phi_2) + \lambda_4 (\Phi_1^\dagger \Phi_2)(\Phi_2^\dagger \Phi_1) \nonumber \\
&&\qquad\qquad + \left \{ \tfrac{1}{2} \lambda_5(\Phi_1^\dagger \Phi_2)^2+ \rm {h.c.}  \right \},
\label{zpot}
\ee
with $\lambda_5$ real.  This corresponds to the ERPS as defined by \eq{erpsgen}.\footnote{After minimizing the scalar potential, one can define the Higgs basis.  It is a simple matter to check that
$Y_1=Y_2$, $Z_1=Z_2$ and $Y_3=Z_6=Z_7=0$, as expected in light of \eq{erps}.}
Minimizing the scalar potential given by \eq{zpot} yields $\vev{\Phi_i^0}=|v_i| e^{i\xi_i}/\sqrt{2}$,
where $\tan\beta\equiv |v_2|/|v_1|$ and $\xi\equiv\xi_2-\xi_1$ depends on the sign of $\lambda_5$.  In particular,
the scalar potential is minimized for $\sin\xi=0$
if $\lambda_5<0$ and $\cos\xi=0$ if $\lambda_5>0$~\cite{Haber:2015pua}.  In the latter case, $\vev{\Phi_2^0}/\vev{\Phi_1^0}=\pm i\tan\beta$.
However, a redefinition of $\Phi_1\to \mp i\Phi_1^0$ yields real non-negative vevs while $\lambda_5\to -\lambda_5$.  
Thus, without loss of generality, we shall assume that the neutral scalar field vevs are real and non-negative, in which case $\lambda_5 \le 0$~\cite{Ginzburg:2010wa}.

\subsection{Yukawa Sector}
\label{sec:Yukawa}

Now we consider the top quark sector.  For now we take the limit where the other SM fermions are massless (it is straightforward to include them as we discuss below). Let $q$ and $u$ denote the $SU(2)$ doublet and singlet top quarks, respectively.  If $q$ and $u$ are invariant under  $\mathbb{Z}_2^m \times\mathbb{Z}_2^i$, then the top Yukawa coupling is forbidden.  In order to allow for the top Yukawa coupling, we include a mirror right-handed top  $U$ that has the same SM gauge quantum numbers as $u$.  We take the top sector to transform under the discrete symmetries as follows,
\be
\mathbb{Z}_2^m:&&  q \Longleftrightarrow q, \quad u \Longleftrightarrow U \nonumber \\
\mathbb{Z}_2^i:&&  q \Longleftrightarrow q,  \quad u \Longleftrightarrow u, \quad U \Longleftrightarrow -U.
\ee
Other choices are possible, including introducing a mirror isospin doublet for $q$ instead of a singlet for $u$, or both. A more sophisticated example is given in Ref.~\cite{Pierce:2007ut}, which has the same Higgs sector but two mirror symmetries acting in the fermion sector differently on the $SU(2)$ doublets and singlets. In the model of Ref.~\cite{Pierce:2007ut}, the second mirror symmetry permits moderately better tuning properties. We employ a right-handed mirror sector in this paper to illustrate the physics with minimal additional field content.

With $U$, the Yukawa couplings take the form,
\be \label{eq:TopYukawa}
V \, \supset \, y_t \left(q \Phi_2 u + q \Phi_1 U \right) + \rm{h.c.}
\ee
In order to avoid anomalies as well as experimental limits on a chiral fourth generation, we introduce a field $\bar U$ in the conjugate representations to $U$ under the SM gauge symmetries. We assign $\bar U$ the following transformations under the discrete symmetries,
\be
\mathbb{Z}_2^m:&&   \bar U \Longleftrightarrow \bar U \nonumber \\
\mathbb{Z}_2^i: &&   \bar U \Longleftrightarrow - \bar U,
\ee
and include a vectorlike mass,
\be
V \, \supset \,  M_U \,U \bar U + \rm{h.c.}
\ee 
The mass term explicitly breaks $\mathbb{Z}_2^m$ but leaves $\mathbb{Z}_2^i$ unbroken.  Because of the $\mathbb{Z}_2^m$ breaking, quantum corrections spoil the degeneracy of the $\Phi_1$ and $\Phi_2$ masses.  However, this breaking is soft, and therefore $m_{22}^2 - m_{11}^2$ is protected from quadratic sensitivity to the cutoff scale $\Lambda$.  We return to the effects of symmetry breaking momentarily.

It is straightforward to include the other fermions of the SM\@.  
The SU(2) doublet of leptons is denoted by $\ell$.  For the remaining right-handed fermions of the SM, $d$, and $e$, we can add mirror partners $D$ and $E$ which receive a vectorlike mass with conjugate fields $\bar D$ and $\bar E$ (we leave the generation indices implicit).  We assume the transformation properties,
\be
\mathbb{Z}_2^m:&&  d \Longleftrightarrow D, \quad e \Longleftrightarrow E, \quad \bar D \Longleftrightarrow \bar D, \quad \bar E \Longleftrightarrow \bar E \nonumber \\
\mathbb{Z}_2^i: &&  d \Longleftrightarrow d, \quad e \Longleftrightarrow e,\quad  D \Longleftrightarrow -D, \quad  E \Longleftrightarrow -E.
\ee
The Yukawa couplings take the form,
\be\label{eq:BotYukawa}
V  \supset y_b \, \left( q \Phi_2^* d + q \Phi_1^* D \right)  + y_\tau \, \left( \ell \Phi_2^* e + \ell \Phi_1^* E \right),
\ee
and the vectorlike fermion masses are,
\be
V & \supset & M_D D \bar D + M_E E \bar E.
\ee

Each vectorlike mass is a source of soft breaking of the mirror symmetry.  For the second scalar to have a natural electroweak scale mass, the mirror right-handed top quark must have a mass near the electroweak scale, while the other mirror fermions can be much heavier.  The proximity of the vectorlike top mass to the weak scale requires a coincidence between two {\it a priori} unrelated mass scales (analogous to the $\mu$-problem in supersymmetry).  This coincidence can be explained if the vectorlike top mass is tied to the mass of dark matter, as can occur in the inert phase with unbroken $\mathbb{Z}_2^i$, or if the vectorlike top mass is related to the mass of an electric and color neutral fermionic state (such as the neutral component of a vectorlike lepton doublet).

For a SM fermion with mass $m_f = y_f v$ and mirror partner with vectorlike mass $M_f$, there is a one-loop correction to $m_{22}^2 - m_{11}^2$.  Requiring this correction to be smaller than the electroweak scale implies a bound of order
\be
M_f \lesssim 4 \pi \frac{v^2}{m_f}.
\ee
The mirror partners of the bottom and tau should be lighter than about 100 TeV, while the partner of the electron should be lighter than about $10^8$~GeV\@. These estimates are robust against inclusion of higher-order corrections because the discrete symmetries imposed on the Higgs fields alone become exact in the limit of small Yukawa couplings. Consequently, only the mirror right-handed top partner needs to have a mass near the electroweak scale, and if the cutoff scale $\Lambda$ is low enough, the other mirror fermions need not be present in the spectrum at all.

Below the scale of the top partner, all the new fermions decouple, and the effective theory is that of the Type I 2HDM~\cite{Haber:1978jt,Hall:1981bc} with approximate discrete symmetries in the Higgs sector. 

\subsection{Discrete Symmetry Breaking Effects}
\label{sec:BreakDiscrete}
As noted previously, the vectorlike masses break part of the discrete symmetry group. We imagine that above some cutoff scale $\Lambda$ the symmetry is restored, while below $\Lambda$, explicit $\mathbb{Z}_2^m$-breaking enters with a characteristic scale $M_U$ which we take to be the vectorlike quark mass. In general there will also be an explicit $\mathbb{Z}_2^m$-breaking mass splitting $m_{11}^2-m_{22}^2$. In the infrared, the splitting is approximately 
\be
\Delta m^2 \equiv m_{22}^2-m_{11}^2\sim  \kappa M_U^2-\frac{3y_t^2M_U^2}{4\pi^2}\log(\Lambda/M_U)\;.
\label{eq:Deltam2}
\ee
Here the first term represents the boundary condition for the splitting at $\Lambda$ ($\kappa$ is a dimensionless coupling), and the second term represents a radiative correction from a quark loop below $\Lambda$. We will have in mind the case where either the $\mathbb{Z}_2^m$ breaking is not communicated at tree-level to the Higgs sector, or $\kappa$ is a somewhat small coupling. In either case, $\Delta m^2$ can be somewhat smaller than $M_U^2$ without fine-tuning, but not much smaller than a loop factor below $M_U^2$.

Because $\mathbb{Z}_2^i$ is unbroken except possibly spontaneously by $v_2$, $m_{12}^2$ is not generated at zeroth order in $v$. Below $M_U$, the fermionic partners can be integrated out, and a splitting between $\lambda_1$ and $\lambda_2$ is generated, since only $\lambda_1$ feels the top quark. The splitting in the infrared (IR) is approximately
\be
\Delta \lambda \equiv \lambda_{1}-\lambda_{2}\sim \frac{3y_t^4}{4\pi^2}\log(M_U/m_t)\sim0.1
\ee
for $M_U\sim1$ TeV\@. The couplings $\lambda_6$ and $\lambda_7$ are not generated due to the unbroken $\mathbb{Z}_2^i$.

At the scale $m_t$ we write $m_{11}^2$ and $m_{22}^2$ in terms of $m^2\equiv\frac{1}{2}(m_{11}^2+m_{22}^2)$ and $\Delta m^2$.
For simplicity, we neglect the effects of $\Delta\lambda$, which are small.

\subsection{Vacua}
\label{sec:vacua}

We define the variables,
\be
\lambda_{345} = \lambda_3 + \lambda_4 + \lambda_5 \qquad R = \frac{\lambda_{345}}{\lambda}\;.
\ee
To prevent a runaway we require~\cite{Ginzburg:2010wa},
\be
\lambda > 0 \qquad R > -1\;.
\label{eq:vacconstr}
\ee
There are different possible electroweak symmetry breaking phases, depending on whether one of the $v_i$ vanishes.

\subsubsection{Inert Phase}

If one of the $v_i$ vanishes, one of the discrete symmetries is unbroken.   We focus on the case where $v_1=0$, in which case the $\mathbb{Z}_2^i$ symmetry is unbroken. This phase is known as the Inert Doublet Model~\cite{Ma:2006km, Barbieri:2006dq}.  At the minimum of the scalar potential, the scalar field vevs are given by\footnote{In the Higgs basis, we identify the Higgs basis field $H_1=\Phi_2$.}
\be
\left< \Phi_2 \right>^2  = v^2 = -\frac{m^2 + \frac{1}{2}\Delta m^2}{\lambda} \qquad \left< \Phi_1 \right> = 0.
\ee
The existence and convexity of the extremum requires that
\begin{align}
\Delta m^2 < -2m^2\;,\;\;\;\;\Delta m^2 < 2m^2\left(\frac{1-R}{1+R}\right)\;.
\end{align}

\subsubsection{Mixed Phase}
There is also a ``mixed phase" where both $v_i\neq0$ and $\mathbb{Z}_2^i$ is spontaneously broken.  Minimizing the scalar potential yields
\begin{align}
m^2&=-\frac{1}{4}\lambda(1+R)v^2\;,\nonumber\\
\tan\beta&=\sqrt{\frac{1-\epsilon}{1+\epsilon}}\;,
\end{align}
where
\begin{align}
\epsilon\equiv\frac{2\Delta m^2}{\lambda(1-R)v^2}\;.
\end{align}
The positivity of $v_{1}^2$ and $v_{2}^2$ and the curvature at the extremum requires
\begin{align}
|R|<1\;,\;\;\;\;\;|\epsilon|<1\;.
\end{align}
Given the constraint on $R$, the constraint on $\epsilon$ can also be written
\begin{align}
m^2<0\;,\;\;\;\;\Delta m^2<-2m^2\left(\frac{1-R}{1+R}\right)\;.
\end{align}
The inert phase can exist without a mixed phase, or vice versa. For both to exist requires 
\begin{align}
|R|<1\;,\;\;\;\;m^2<0\;,\;\;\;\;\Delta m^2 < 2m^2\left(\frac{1-R}{1+R}\right)\;.
\end{align}
The energy densities at the inert and mixed extrema are
\begin{align}
V_I=-\,\frac{(2m^2+\Delta m^2)^2}{8\lambda}\;,\;\;\;\;V_M=-\,\frac{1}{4\lambda}\left(\frac{(2 m^2)^2}{1+R}+\frac{(\Delta m^2)^2}{1-R}\right)\;.
\end{align}
Taking the difference, we can see that the mixed phase minimum is deeper than the inert minimum when both exist.

\subsection{Scalar Spectrum}
\label{sec:spec}
\subsubsection{Inert Phase}
In the inert phase, the neutral CP-even scalar masses squared are
\begin{align}
m_h^2&=\lambda v^2\;,\nonumber\\
m_H^2&=-\frac{1}{2}\lambda v^2(1-R)-\Delta m^2\;.
\end{align}
Since $\Delta m^2$ can be arbitrarily negative in the inert phase (for example, $M_Q=M_U\gg v$), in principle $m_H^2$ can be arbitrarily heavy. Of course, we are interested in the case where it sits near the electroweak scale.

The other Higgs boson masses are given by
\begin{align}
m_A^2&=m_H^2-\lambda_5 v^2\;,\nonumber\\
m_{H^\pm}^2&=m_H^2-\frac{1}{2}(\lambda_4+\lambda_5)v^2\;.
\end{align}
Since $\lambda_5\leq0$, the CP-odd scalar is heavier than the CP-even scalar.   If $\lambda_4\geq\lambda_5$, then $H$ can be the lightest electrically neutral, $\mathbb{Z}_2^i$-charged particle, as required for $H$ dark matter.

\subsubsection{Mixed Phase}
In the mixed phase, the neutral CP-even scalar squared-mass matrix is
\begin{align}
\mathcal{M}_e^2=\frac{1}{2}\lambda v^2\left(
\begin{array}{cc}
 1+\epsilon & R\sqrt{1-\epsilon^2}    \\[8pt]
 R\sqrt{1-\epsilon^2} & 1-\epsilon
\end{array}
\right)\;.
\label{eq:Mh2e}
\end{align}
Fits to Higgs coupling measurements indicate that if a 2HDM is realized in nature, it must lie near an alignment limit~\cite{Gunion:2002zf,Craig:2012vn,Craig:2013hca,Haber:2013mia,Carena:2013ooa}, where the CP-even scalar mixing angle $\alpha\simeq\beta-\pi/2$. In this limit, one of the states is ``mostly aligned" with the vev $(v_1,v_2)$, and therefore has SM-like couplings to electroweak gauge bosons, so it can be identified with the 125 GeV Higgs boson. 

The mixing angle and spectrum analysis is most convenient in the Higgs basis, where the squared-mass matrix given in Eq.~(\ref{eq:Mh2e}) is rotated by an angle $\beta$. In this basis, 
\begin{align}
\mathcal{M}_e^2=\frac{1}{2}\lambda v^2\left(
\begin{array}{cc}
 (1+R)+(1-R)\epsilon^2 & (1-R)\epsilon\sqrt{1-\epsilon^2}    \\[8pt]
 (1-R)\epsilon\sqrt{1-\epsilon^2}& (1-R)(1-\epsilon^2 )
\end{array}
\right)\;.
\label{eq:Mh2eHB}
\end{align}
In this basis, the mixing angle is $\alpha-\beta+\pi/2$. If $|\cos(\beta-\alpha)|$ is small, the approximate eigenvector $(1,0)$ is aligned with the vev and is SM-like. It will be the lightest scalar state if the first diagonal element is smaller than the second, 
\begin{align}
 (1+R)+(1-R)\epsilon^2 < (1-R)(1-\epsilon^2 )\;,
\end{align}
which, along with the vacuum constraint Eq.~(\ref{eq:vacconstr}), implies that $R$ should lie in the range
\begin{align}
-1<R<-\frac{\epsilon^2}{1-\epsilon^2}\;.
\label{eq:Rbound}
\end{align}

\begin{figure}[t]
\begin{center}
\includegraphics[scale=0.75]{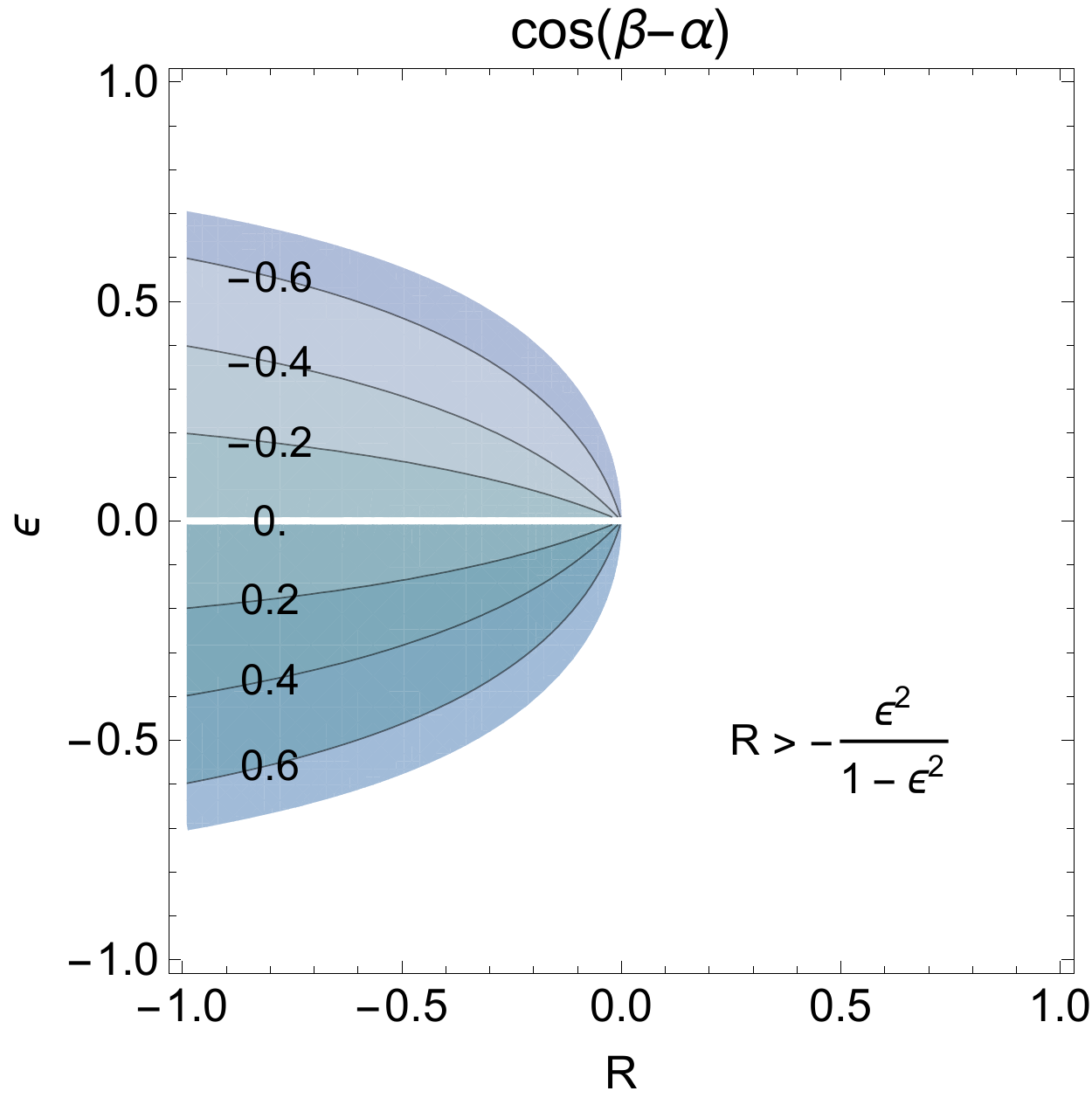}
\end{center}
\caption{The alignment limit parameter $\cos(\beta-\alpha)$. We mask regions where the heavier CP-even state is the more SM-like of the two.}
\label{fig:cba}
\end{figure}

\begin{figure}[t]
\begin{center}
\includegraphics[scale=0.75]{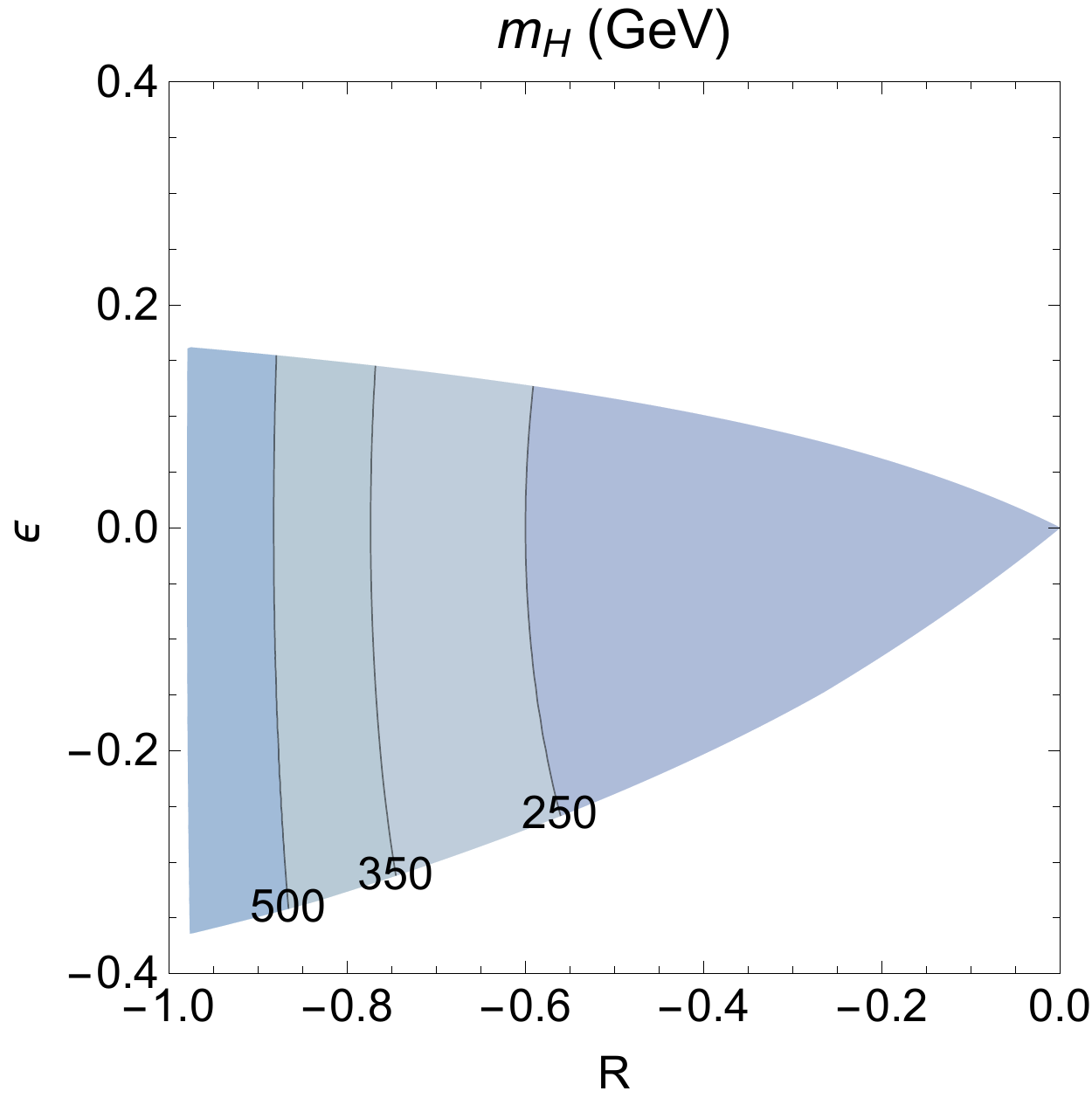}
\end{center}
\caption{The heavy CP-even scalar mass $m_H$ (parameter space zoomed in compared to Fig.~\ref{fig:cba}). We mask regions where the alignment parameter is in tension with Higgs boson coupling fits (taken from~\cite{Craig:2015jba}), and regions where $\lambda > 8\pi$ (corresponding to $R\lesssim -0.98$).}
\label{fig:mH}
\end{figure}

The smallness of the mixing angle requires that the off-diagonal element of (\ref{eq:Mh2eHB}) be small compared to the splitting of the diagonal elements. The off-diagonal element vanishes for $R=1$, $\epsilon=\pm1$, and $\epsilon=0$, but it is easy to see that only in the case $\epsilon=0$ can the SM-like state be the lightest, and then only if Eq.~(\ref{eq:Rbound}) is satisfied, implying $R<0$.  When $\epsilon$ and $\cos(\beta-\alpha)$ are both small and nonzero, the latter is approximately given by
\begin{align}
\cos(\beta-\alpha)\simeq-\frac{\epsilon(1-R)}{2|R|}\;.
\end{align}
Self-consistency of this result requires $|\epsilon/R|<1$ (since we have already eliminated the possibility that $R\approx 1$ when $\epsilon$ is small.) In Fig.~\ref{fig:cba} we plot the exact $\cos(\beta-\alpha)$ on the $(R,\epsilon)$ plane, and see that as expected, alignment favors small $|\epsilon|$.

The CP-even scalar squared-masses are given by
\begin{align}
m_{h,H}^2&=\frac{1}{2} \lambda v^2(1\mp\sqrt{R^2+(1-R^2)\epsilon^2})\;,
\end{align}
and the other Higgs masses are given by
\beq
m_A^2=-\lambda_5 v^2\;,\qquad\quad
m_{H^\pm}^2=-\frac{1}{2}(\lambda_4+\lambda_5)v^2\;.
\eeq
We can solve for $\lambda$ in terms of $m_h$. Inserting the solution for $\lambda$ into the heavy neutral CP-even scalar mass, we obtain an expression for $m_H^2$ as a function of $R$ and $\epsilon$:
\begin{align}
m_H^2=m_h^2\left(\frac{1+\sqrt{R^2+(1-R^2)\epsilon^2}}{1-\sqrt{R^2+(1-R^2)\epsilon^2}}\right)\;.
\end{align}
The heavy scalar mass is plotted in Fig.~\ref{fig:mH} with $m_h=125$ GeV\@. Its growth as 
$R\to -1$ 
is due to the fact that in that limit, holding $m_h^2$ fixed requires increasing $\lambda$ as $(1+R)^{-1}$. Conversely, 
the vanishing of $m_{12}^2$ as imposed by the exact $\mathbb{Z}_2^i$ symmetry prevents the decoupling of $m_H$ while keeping dimensionless couplings finite. 
The quartic $\lambda$ reaches the unitarity limit of $8\pi$ for $R\approx -0.98$.

In Fig.~\ref{fig:mH} we crop regions where the alignment parameter is in tension with Higgs coupling fits~\cite{Craig:2015jba}. Positive $\epsilon$ corresponds to negative $\cos(\beta-\alpha)$ and $\tan\beta<1$, which is where the coupling constraints are strongest, while negative $\epsilon$ corresponds to the more weakly-constrained region of positive $\cos(\beta-\alpha)$ and 
$\tan\beta>1$; these differences are responsible for the asymmetry in the allowed region.\footnote{The relative strength of the constraints can be understood as follows. The fermionic couplings are universally scaled by $\sin(\beta-\alpha)+\cos(\beta-\alpha)/\tan\beta$, and for $\tan\beta\sim$ few, as $|\cos(\beta-\alpha)|$ increases to moderate values $\sim 0.3$, this scaling factor eventually provides a suppression to the couplings for either sign of $\cos(\beta-\alpha)$.  However, the suppression is more severe for negative $\cos(\beta-\alpha)$ than for positive, so the constraints are stronger for the former. In the region of $\tan\beta<1$, for small $\cos(\beta-\alpha)$, there is a rapid enhancement towards positive $\cos(\beta-\alpha)$ and a rapid suppression towards negative $\cos(\beta-\alpha)$, so the parameters quickly become disallowed. However, a 50\% decrease in the coupling gives a bigger change in the rates than a 50\% increase in the coupling, so again the constraints are somewhat stronger for negative $\cos(\beta-\alpha)$. }

\begin{figure}[t!]
\begin{center}
\includegraphics[scale=0.75]{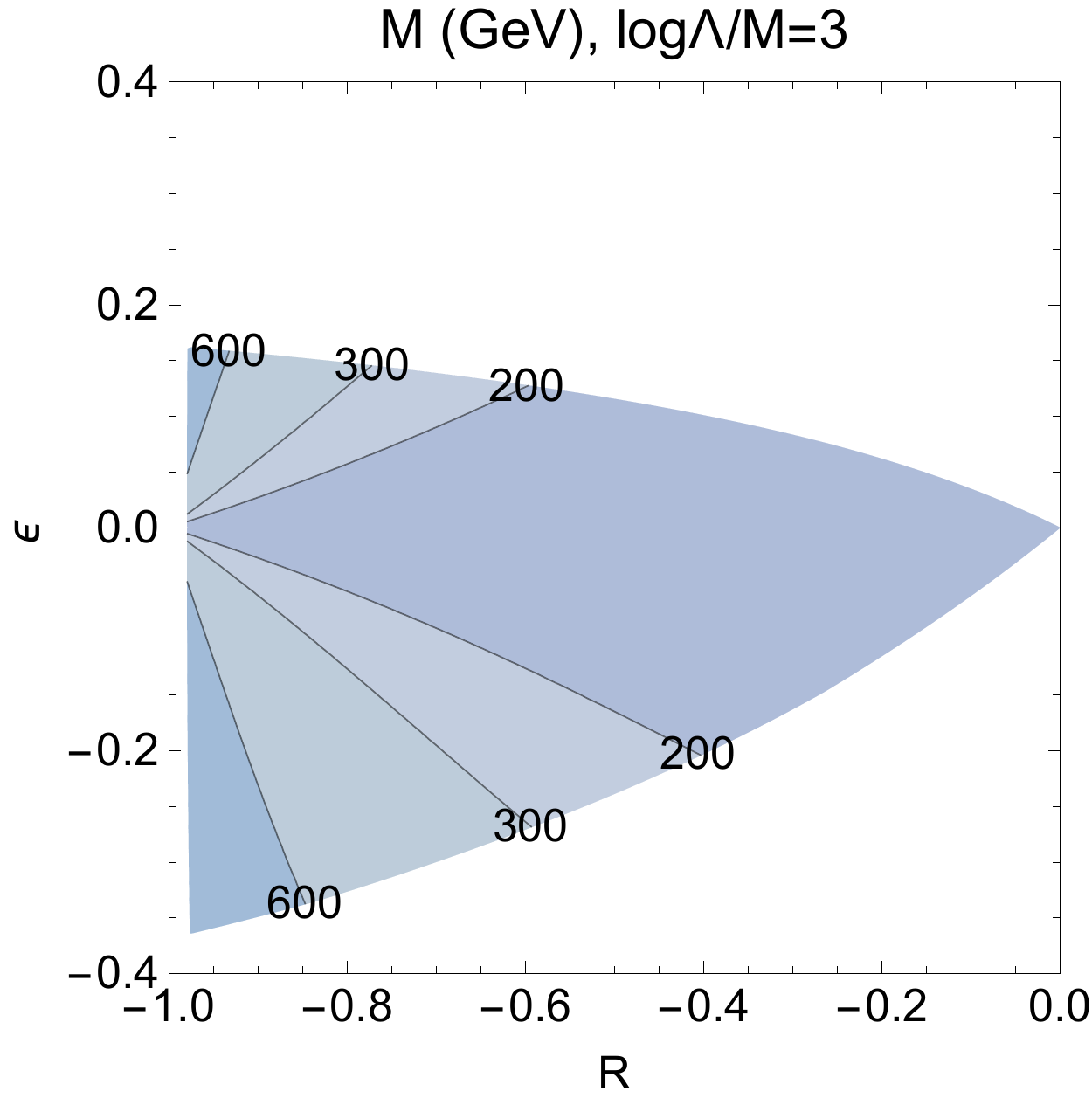}
\end{center}
\caption{The vectorlike mass scale $M$ (parameter space zoomed in compared to Fig.~\ref{fig:cba}). We mask regions as in Fig.~\ref{fig:mH}.}
\label{fig:Mlog3}
\end{figure}

Without a cancellation between the terms in Eq.~(\ref{eq:Deltam2}), $\Delta m^2$ is expected to be at least as large as the logarithmic correction.  Using 
\begin{align}
\Delta m^2 = m_h^2\left(\frac{\epsilon(1-R)}{1-\sqrt{R^2+\epsilon^2(1-R^2)}}\right)\;
\end{align}
and setting $\Delta m^2$ equal to the logarithmic term in Eq.~(\ref{eq:Deltam2}), we obtain $M$ shown in Fig.~\ref{fig:Mlog3}  as a function of $R$, $\epsilon$, and the cutoff $\Lambda$,
under the assumption that $\log(\Lambda/M)=3$. 
We see that low vectorlike masses are obtained unless $R$ is close to $-1$.

\begin{figure}[t!]
\begin{center}
\includegraphics[scale=0.75]{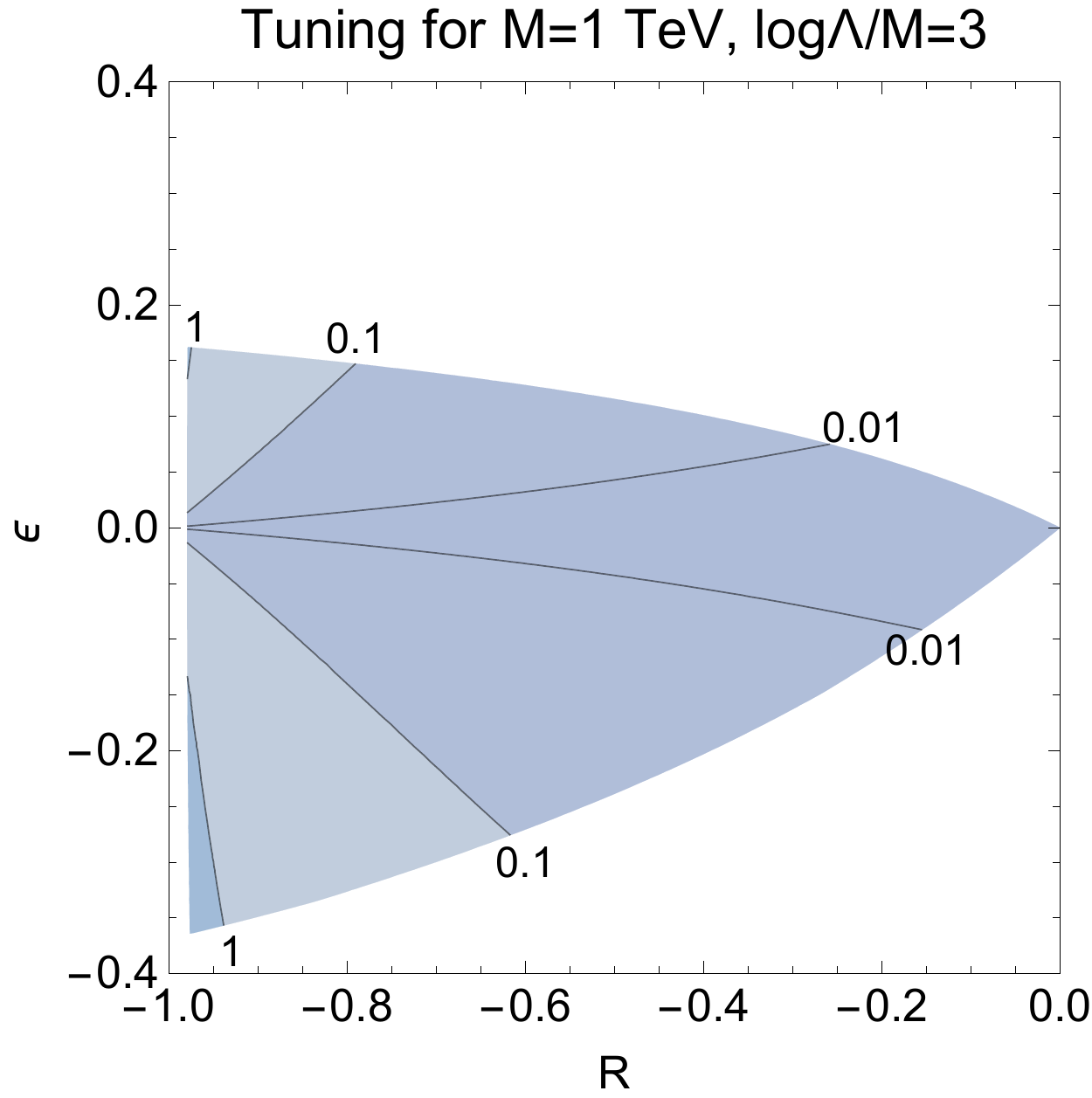}
\end{center}
\caption{The percent cancellation between the tree-level and loop-induced contributions to $\Delta m^2$ in Eq.~(\ref{eq:Deltam2}) needed to raise the scale of the vectorlike fermions to 1 TeV\@. We mask regions as in Fig.~\ref{fig:mH}.}
\label{fig:M1000Tuning}
\end{figure}

Such low values for $M$ are in tension with direct search limits at the LHC\@, as discussed in section \ref{sec:pheno}. However, there are situations in which the vectorlike scale might be higher. For example, a part-in-$N$ tuning in Eq.~(\ref{eq:Deltam2}) allows raising the scale $M$ by about a factor of $\sqrt{N}$. In Fig.~\ref{fig:M1000Tuning} we plot the percent cancellation between the terms in Eq.~(\ref{eq:Deltam2}) necessary to raise $M$ to 1 TeV, and find that the tuning is modest, of order 10\%, in much of the allowed parameter space.

Another interesting possibility for raising $M$ is the introduction of soft $\mathbb{Z}_2^i$-breaking through the $m_{12}^2$ parameter. In order to avoid an unnatural coincidence of scales, it would be desirable to break all of the discrete symmetries with the same spurion. We leave further analysis of this scenario for future work.

\subsection{Phenomenology}
\label{sec:pheno}

Our scenario consists of a 2HDM augmented by vectorlike quarks.  The (partially) natural parameter space of our model predicts electroweak scale masses for the  vectorlike quarks and extra Higgs bosons, leading to signals at colliders such as the LHC\@. The collider signals  depend on the phase of the vacuum.  First, we discuss the inert phase, where $\mathbb{Z}_2^i$ is preserved.  Then we consider the phenomenology of the mixed phase, where $\mathbb{Z}_2^i$ is spontaneously broken.

In the inert phase, the lightest odd state under $\mathbb{Z}_2^i$ is collider stable.  If the lightest odd state is the vectorlike quark, $U$, then searches for Heavy Stable Charged Particles (HSCPs) require $m_U \gtrsim 1.3~\mathrm{TeV}$~\cite{CMS:2015kdx}.  Similarly, if the lightest odd state is the charged Higgs, $H^\pm$, then HSCP searches require $m_{H^{\pm}} \gtrsim 400~\mathrm{GeV}$~\cite{Chatrchyan:2013oca}.  If the lightest odd state is the neutral Higgs, $H$, then it is a dark matter candidate~\cite{Pierce:2007ut} and it contributes to missing energy at colliders. In this case, the vectorlike quark can decay to Higgs bosons, $U \rightarrow (t H, t A, b H^\pm)$.  This leads to signals that are similar to stop production in supersymmetry, and $U$ is constrained by searches for stops.
If $\mathrm{Br}(U \rightarrow t H) = 100\%$, searches for $t \bar t$ plus missing energy constrain $m_U \gtrsim 900$~GeV for $m_H\lesssim 400$~GeV~\cite{Khachatryan:2016oia}.\footnote{In order to interpret the limit on $U$ from stop searches, we have assumed similar efficiency for $U$ and stops.  This is only approximately true due to different production kinematics for colored scalars versus fermions.  Dedicated searches for fermionic top partners in the $t \bar t$ plus missing energy final state have not been updated since $\sqrt s = 7$~TeV, but require $m_U\gtrsim 500$ GeV for $\mathrm{Br}(U \rightarrow t H) = 100\%$ and $m_H\lesssim 150$ GeV~\cite{Aad:2012uu}.}
 If $\mathrm{Br}(U \rightarrow t H) = 25\%$, we require $m_U \gtrsim 650~\mathrm{GeV}$ for $m_H\lesssim 200$~GeV~\cite{Khachatryan:2016oia}.  There are also constraints on $U \rightarrow b (H^\pm \rightarrow {W^\pm}^* H)$ from searches for stop decays to charginos.  For example, if $\mathrm{Br}(U \rightarrow b H^{\pm}) = 100~(50) \%$, then we require $m_U \gtrsim 800~(700)~\mathrm{GeV}$, for $m_{H^\pm} \lesssim 400~(200)~\mathrm{GeV}$, assuming $m_{H^\pm} \approx 1/2 \, (m_U + m_H)$~\cite{Khachatryan:2016oia}.  Searches at the LHC for electroweak production of heavy stable Higgs bosons in association with a monojet will probe masses up to of order 1 TeV~\cite{Harris:2014hga,Haisch:2015ioa}.

In the mixed phase, the $U$ quark mixes with the top quark through the $y_t$ coupling of Eq.~(\ref{eq:TopYukawa}), and can therefore decay also to $Wb$, $Zt$, and $ht$.  If the sum of these branching ratios is 100\%, then experimental constraints require $m_U \gtrsim 700-900$ GeV (depending on the relative branching ratios)~\cite{Aad:2015kqa,Khachatryan:2015oba,ATLAS-CONF-2016-013}. In the model discussed here the $tH$ and $bH^+$ modes are likely to dominate when they are kinematically accessible since the couplings are unsuppressed by mixing. Mixing also allows $H$ to decay to SM particles, with dominant branching ratios into $t\bar t$ $(b\bar b)$ above (below) the top pair threshold. The $H^+$ branching ratios depend on whether the $HW^+$ and $tb$ modes are kinematically accessible. LHC searches for $t\bar t H$, $H\rightarrow t\bar t$ and $\bar{t}bH^+$, $H^+\rightarrow t\bar b$ offer promising reach for direct production of new scalar states near the alignment limit~\cite{Craig:2015jba,Hajer:2015gka}.

\section{Supersymmetry}
\label{sec:SUSY}
Supersymmetry is a highly motivated 2HDM, and if the soft SUSY-breaking scale is low, then both doublets are naturally light. On the other hand, there are reasons to contemplate a high SUSY-breaking scale. For example, while the MSSM can accommodate a 125 GeV Higgs, it requires dimensionful parameters (either soft masses or $A$-terms) significantly above the electroweak scale in the stop sector\cite{Draper:2016pys}. In such models a soft Higgs mass must be fine-tuned in order to obtain $v/m_{soft}\ll1$.

 Without additional tuning or symmetries, the other Higgs states will be as heavy as the squarks, usually completely decoupled and too heavy to produce at the LHC\@.  This can be seen as follows. Given the MSSM Higgs sector input parameters $m_{H_u}^2$, $m_{H_d}^2$, $b$, and $\mu$, the light electroweak scale requires tuning $m_{H_u}^2$,
\begin{align}
m_{H_u}^2\approx -\mu^2+\frac{b^2}{\mu^2+m_{H_d}^2}\;.
\end{align}
While it is technically natural for $b$ and $\mu$ to lie anywhere between the electroweak scale and $m_{soft}$, $m_{H_d}^2$ is not generically protected by symmetries, and sets the scale of the other Higgs multiplets to be of order $m_{soft}$ via
\begin{align}
m_A^2\approx \frac{b^2}{\mu^2+m_{H_d}^2}+\mu^2+m_{H_d}^2\approx m_{soft}^2\;.
\end{align}
Discrete symmetries of the type discussed in the previous section may be imposed among the soft masses at the SUSY-breaking mediation scale in order to naturally tie the Higgs doublet masses together. Without adding chiral multiplets to the MSSM, these symmetries are again typically broken by dimensionless Yukawa couplings, leading to splittings controlled by a product of Yukawa couplings and soft masses.

There are several ways to proceed. First, we can attempt to introduce mirror vectorlike multiplets such that the Yukawa terms in the superpotential preserve a mirror symmetry. We find that the model and the mirror symmetry discussed previously cannot be generalized in a simple way consistent with the gauge interactions and supersymmetry if $H_u$ and $H_d$ are required to be present at low energies. A second possibility we consider is a weaker but more minimal mechanism, requiring a top-bottom symmetry but no new states; in this case there are hard discrete symmetry breaking effects that must be made as small as possible.

\subsection{C and CP-symmetric SSMs}
We promote all fields to superfields. As usual, holomorphy of the superpotential requires that two chiral Higgs superfields carry opposite hypercharge. Therefore, the most straightforward implementation of the discussion in the previous sections is to add both right-handed top partners and {\it two} mirror doublets $H_u'$ and $H_d'$. The Higgs sector is now a 4HDM with a quadratic scalar potential of the form
\begin{align}
V\supset m_1^2(|H_d|^2+|H_d'|^2)+m_2^2(|H_u|^2+|H_u'|^2)-m_3^2(H_uH_d+H_u'H_d')\;.
\end{align} 
Since $m_1^2$ and $m_2^2$ are still independent, two linear combinations of fields can be made light with one tuning, while the other two linear combinations remain heavy.\footnote{In principle, we could achieve two light fields and one heavy field in a supersymmetric 3HDM, for example by omitting $H_d'$ and making $H_d$ invariant under the mirror symmetry. This model would have anomalies and no $\mu$ term, so we stick with the case where even the heavy doublets have mirror partners.} If, for example, $m_1^2$ is not tuned and remains of order $m_1^2\sim m_{soft}^2\gg v^2$, we can integrate out $H_d$ and $H_d'$,
\begin{align}
H_d\approx\frac{m_3^2}{m_1^2}H_u^*\;,\;\;\;H_d'\approx\frac{m_3^2}{m_1^2}H_u'^*\;,
\end{align}
and the low energy theory is just a particular case of the 2HDM studied in the previous section. This is analogous to the ``decoupling limit" of the MSSM\@.

A more interesting question is whether we can implement sufficient discrete symmetries on the minimal SUSY Higgs sector (i.e. without the introduction of $H_u'$ and $H_d'$) so that $H_u$ and $H_d$ are light fields. Identifying $\Phi_1$ and $\Phi_2$ in the common notation as $H_d$ and $H_u$, respectively, the hypercharges of $\Phi_1$ and $\Phi_2$ are opposite. Thus the mirror symmetry in Eq.~(\ref{eq:mirrorsym}) must become a charge-conjugation symmetry. The superpotential must be modified from Eqs.~(\ref{eq:TopYukawa}) and (\ref{eq:BotYukawa}); one possibility is
\begin{align}\label{eq:W}
W \, \supset &\, y_t \left(q \Phi_2 u + \bar Q \Phi_1 \bar U \right)+y_b \, \left( q \Phi_1 d + \bar Q \Phi_2 \bar D \right) + y_\tau \, \left( \ell \Phi_1 e + \bar L \Phi_2 \bar E \right) \nonumber\\
&+ M_Q \,Q \bar Q + M_U \,U \bar U +M_D D \bar D + M_L L \bar L + M_E E \bar E\;.
\end{align}
We have introduced mirror left-handed fields and their conjugates $Q, \bar Q,\dots$ in addition to the right-handed mirrors $U, \bar U,\dots$ in order to write gauge invariant supersymmetric Yukawa interactions and vectorlike mass terms. For brevity we have only written a subset of the possible mass terms. If, as before, we enforce the $\mathbb{Z}_2^m$ on the Yukawa interactions, but allow soft breaking by the vectorlike masses, the mirror symmetry can be defined in a variety of ways. One assignment is
\be
\mathbb{Z}_2^m:&&  q \Longleftrightarrow \bar Q, \qquad u \Longleftrightarrow \bar U, \qquad d \Longleftrightarrow \bar D, \qquad \ell \Longleftrightarrow \bar L,  \qquad e \Longleftrightarrow \bar E,  \nonumber  \\
&& Q \Longleftrightarrow  Q^*, \quad\,  U \Longleftrightarrow  U^*, \quad\,\,  D \Longleftrightarrow  D^*, \quad\,  L \Longleftrightarrow L^*, \quad\,\, E \Longleftrightarrow E^*\;.
\ee
Here $\mathbb{Z}_2^m$ is acting as a generalization of the ordinary charge conjugation on $q,\bar Q,u,\bar U,d,\bar D,\dots$ and like CP on the $Q,U,D,\dots$\footnote{For simplicity we are ignoring small explicit CP-violation in the Yukawa couplings; this will turn out not to matter.} The difference is problematic because it means the gauge interactions cannot be completely invariant: the C and CP transformations of the vector superfields are different, so the $Q$ and $\bar Q$ gauge interactions are not both preserved. Similarly, gauge interactions break the symmetries if $Q,U,D,\dots$ are defined to be invariant under $\mathbb{Z}_2^m$. Since these are hard breakings, in general the symmetries cannot be protected at low energies.

The C symmetry can be promoted to generalized CP, e.g., $u \Longleftrightarrow U^*$ and $\bar U\Longleftrightarrow \bar U^*$. However, this is not a symmetry of the superpotential, and moreover it cannot be a mirror symmetry and a symmetry of the Yukawa sector unless
$\Phi_2$ has the same hypercharge as $\Phi_1$, in which case the down-type Yukawa couplings are forbidden. 
Alternatively, we can impose $\mathbb{Z}_2^m$ as a P symmetry on the Higgs fields $\Phi_2\Longleftrightarrow\Phi_1^*$, and define $u\Longleftrightarrow \bar U^*$, etc. Once again, however, there is no transformation for the unbarred mirror fields that preserves the gauge interactions.

Ref.~\cite{Pierce:2007ut} considered a different pattern of discrete symmetries which includes two mirror symmetries. It is similarly not possible to extend their model to a supersymmetric one in a minimal way. The $SU(2)$ doublet fermions are invariant under one of the mirror symmetries and the $SU(2)$ singlets are invariant under the other, implying neither can be preserved with generalized C and P transformations in supersymmetry.

\subsection{$t/b$--symmetric MSSM}
Hard discrete symmetry breaking signals that radiative corrections will generate mass splitting proportional to the typical soft mass scale. Whether or not the splitting is large compared to the electroweak scale is then a detailed model-dependent question of the size of the coefficient. This is a weaker position than in the non-supersymmetric case where we avoided hard breakings completely. However, it can be an interesting possibility to consider precisely because it strongly constrains the allowed UV completions. 

With this in mind we could analyze the radiative splittings generated by the hard breaking terms in the mirror model of the previous section. However, perhaps a more interesting model is simply the MSSM without mirror matter. We study this possibility in this section. Explicit symmetry breaking from hypercharge and Yukawa interactions will limit the parameters and UV completions that naturally contain a light second doublet.

Unsurprisingly, having both doublets light requires the MSSM Higgs sector mass parameters to be of order the electroweak scale. To see this, note that the neutral Higgs masses satisfy the tree-level relations
\beq
m_h^2+m_H^2=m_A^2+m_Z^2\,,\qquad\qquad\quad
m_A^2=-m_{12}^2(\tan\beta+\cot\beta)\;,
\eeq
while the minimization conditions take the form
\beq
m_1^2=-m_{12}^2\tan\beta+\mathcal{O}(v^2)\;,\qquad\qquad
m_2^2=-m_{12}^2\cot\beta+\mathcal{O}(v^2)\;.
\eeq
Therefore, having both $m_h$ and $m_H$ of order the electroweak scale requires $-m_{12}^2\tan\beta$ to be of order $v^2$, which in turn requires $m_1^2$ and $m_2^2$ to be of order $v^2$.

Small $m_{12}^2$ is radiatively stable. The beta function for $m_{12}^2$ contains terms of order $m_{12}^2$ and $\mu M_i$, where $M_i$ is an electroweakino mass. If $\mu$ and $m_{12}^2$ are small at the mediation scale $\Lambda$ (consistent with the discrete symmetries), they remain small under renormalization group (RG) evolution. 

On the other hand, $m_1^2$ and  $m_2^2$ are naturally of order the heaviest sparticle mass. For values of $\tan\beta$ somewhat above 1, the tuning of the electroweak scale is primarily a tuning of $m_2^2$, so naturally obtaining the second light doublet requires $m_1^2$ to remain of order $m_2^2$. At the mediation scale, $m_1^2=m_2^2$ may be a consequence of the approximate discrete symmetries. Under RG evolution, a splitting develops, which at 1-loop is approximately\cite{Martin:1997ns}
\begin{align}
\Delta(m_1^2-m_2^2)\simeq \frac{1}{8\pi^2}\log\left(\frac{\Lambda}{m_{soft}}\right)\left(3Z_t-3Z_b-Z_\tau+\frac{3}{5}g_1^2 D_Y\right)\;,
\end{align}
where
\begin{align}
Z_t&=y_t^2(m_{H_u}^2+m_{Q_3}^2+m_{u_3}^2+|A_t|^2)\;,\\
Z_b&=y_b^2(m_{H_d}^2+m_{Q_3}^2+m_{d_3}^2+|A_b|^2)\;,\\
Z_\tau&=y_\tau^2(m_{H_d}^2+m_{L_3}^2+m_{e_3}^2+| A_\tau |^2)\;,\\
D_Y&=\rm{Tr} (Y m^2)\;.
\end{align}
The hypercharge-generated term $D_Y$ is small in many explicit models at the mediation scale, and this property is radiatively stable because its beta function is homogeneous~\cite{Demir:2004aq,Carena:2010gr}. For $\tan\beta\sim 50$, the role of the mirror matter fields in the exchange symmetry can be partially approximated by a $t\leftrightarrow b$ symmetry in the Yukawa couplings and soft masses (but which is again broken by hypercharge and the $\tau$ Yukawa coupling). At 1-loop this symmetry leads to a cancellation of the $Z_b$ and $Z_t$ terms above. However, there is no candidate exchange partner for the (s)leptons.  If right-handed neutrinos are included to form a Dirac mass with SM neutrinos, then the neutrino Yukawa coupling is much too small.  If Majorana right-handed neutrinos are included, the right-handed multiplet decouples at a high scale or, again, the Yukawa coupling is too small.  Therefore, for small $D_Y$ and large $\tan\beta$, the dominant 1-loop contribution to the $m_1^2-m_2^2$ splitting may come from terms involving the $\tau$ Yukawa coupling, which in this case is non-negligible, of order $y_\tau\sim1/2$. To avoid fine-tuning, if the mediation scale is low, then the stau masses must be at most an order of magnitude above the mass of the second Higgs doublet. If the mediation scale is high, of order the GUT scale, then the staus and Higgs bosons must be of similar masses.

There are also important 2-loop contributions, including effects from the 1-loop $y_b$-$y_t$ splitting induced by hypercharge and $y_\tau$. The dominant effects are roughly of order
\begin{align}
\Delta_2(m_1^2-m_2^2)&\simeq \frac{1}{2} \frac{6y_t^2y_\tau^2}{8\pi^2}\frac{m_{Q_3}^2}{16\pi^2}\log^2(\Lambda/m_{soft})\,,
\end{align}
where $\Lambda$ is the mediation scale.\footnote{Here for simplicity we have used only one boundary scale $\Lambda$, but in principle the ``flavor scale" at which the symmetries fix $y_b=y_t$ may be different.}
Avoiding fine-tuning with a low scale $\Lambda$ implies third-generation squark masses less than two orders of magnitude above the mass of the second Higgs doublet. If the mediation scale is high, the third-generation squarks and Higgs bosons must be of the same order -- but in this case the fine-tuning of the electroweak scale itself is relieved.

In conclusion, it is possible to imagine a ``mini-split"-type scenario~\cite{Arvanitaki:2012ps} with two light Higgs doublets and only one meso-tuning. It is most easily achieved in a framework like gauge mediation, where $m_{H_u}^2=m_{H_d}^2$ at the messenger scale and $D_Y$ vanishes (in both cases, up to possible corrections from couplings generating $\mu$ and $m_{12}^2$). Furthermore in gauge mediation the $m_{u_3}^2$ and $m_{d_3}^2$ boundary values are split only by hypercharge effects, and the messenger scale can be low. In such a scenario the higgsino states are expected to be as light as the extra Higgs states, and staus are also expected to be light, at most an order of magnitude heavier. A gauge mediated model with a mini-split spectrum, low $\tan\beta$, and large $m_A$ was studied in~\cite{Cohen:2015lyp}, and it would be interesting to examine whether variants of the model can accommodate large $\tan\beta$ and low $m_A$. We leave this investigation for future work.

\section*{Acknowledgments}
We thank Nima Arkani-Hamed, Lawrence Hall, Jesse Thaler, and Neal Weiner for helpful conversations.  P.D. is supported in part by the National Science Foundation Grant No.\ PHY13-16748.  H.E.H. is supported in part by U.S. Department of Energy grant DE-FG02-04ER41286. 
H.E.H. is grateful for the hospitality of the Theory Group at the Lawrence Berkeley National Laboratory, where this work was initiated.  H.E.H. and J.T.R. acknowledge the hospitality and the inspiring working atmosphere  
of the Aspen Center for Physics, supported by the National Science Foundation Grant No.\ PHY-1066293, where some of this work was carried out.   This research was completed at
the Kavli Institute for Theoretical Physics in Santa Barbara, CA, and supported in part by the National Science Foundation under Grant No. NSF PHY11-25915.

\end{document}